\documentclass[12pt] {article}
\def\m{\multicolumn}
\newfont{\jnp}{cmss10}
  \def\n{\noindent}
      \def\bea{\begin{eqnarray}}
  \def\eea{\end{eqnarray}}    
  \def\be{\begin{equation}}  \def\ee{\end{equation}}
\begin{document}
{\begin{center}

{\Large\bf Properties of doubly charmed baryons in the quark-diquark model}\\

{\large Ajay Majethiya \footnote{ajay.phy@gmail.com}$^ \dag$, Bhavin Patel$^ \dag$, Ajay Kumar Rai* , and \\ P C Vinodkumar$^ \dag$}\\
{$^ \dag$ \small Department of Physics, Sardar Patel University,\\
  Vallabh Vidyanagar, Anand, Gujarat-388 120.}\\
{\small *Department of Applied Sciences and Humanities, \\
  SVNIT, Surat-395 007, Gujarat, INDIA.}\\
\end{center}}

\abstract{ Baryons containing two heavy quarks are important and interesting systems to
study the quark-diquark structure of baryons and to understand the dynamics of QCD at
hadronic scale. The Selex Collaboration has recently reported the discovery of
$\Xi_{cc}$ with a mass of $3.519$ Ge$V$. However, other groups such as Babar , Belle
 and Focus have all failed to confirm this
state. So there is a demand to review this state both theoretically and experimentally.
We report here the spectra and magnetic moments of $ccq(q\in u,d,s)$ systems using
coulomb plus Martin potential. Here the two heavy quarks are considered for the diquark
states. The same potential form is used for the diquark interaction as well as for the
quark-diquark interaction. The chromomagnetic one gluon exchange interaction are
perturbatively treated here to get the masses of $J^P={\frac{1}{2}}^{+}$ and
$J^P={\frac{3}{2}}^{+}$ states. Accordingly, we obtain $\Xi^{++}_{cc}$ (3519, 3555),
$\Xi^{+}_{cc}$ (3527, 3562) and $\Omega^{+}_{cc}$ (3648,3688) states of the $ccq$
$(\frac{1}{2}^{+},\frac{3}{2}^{+})$ combinations. Our predictions are in accordance
with the selex result for $\Xi^{++}_{cc}$ (3519) and other model predictions.
Predictions of other properties of these $ccq$ systems will be presented in detail.
}\\

 \noindent
{\large{\bf Introduction}}\\
{\Large} The first observation of $B_{c}^+$ meson by the CDF collaboration \cite{F.
Abe1998} opens a new direction in the physics containing heavy quarks. This particle
completes the list of heavy $Q\bar Q$ mesons accessible for the experimental
investigations. It begins another list of hadrons containing two heavy quarks. The
investigation of properties of hadrons containing heavy quarks is of great interest in
understanding the dynamics of QCD at the hadronic scale. Though the experimental and
theoretical data on the properties of heavy flavour mesons are available plenty in
literature, the masses of most of the heavy baryons have not been measured yet
experimentally \cite{arXiv:hep2000}. Thus the recent predictions about the heavy baryon
mass spectrum have become a subject of renewed interest due to the experimental
facilities at Belle, BABAR, DELPHI, CLEO, CDF etc
\cite{Mizuk2005,Aubert2007,Feindt2007,Edwards1995,Artuso2001,I.V.Gorelov2007}. These
experimental groups have been successful in discovering heavy baryonic states along
with other heavy flavour mesonic states and it is expected that more heavy flavour
baryon states will be detected in near future. Most of the new states are within the
heavy flavour sector with one or more heavy flavour content and some of them are far
from most of the theoretical predictions. Though there are consensus among the
theoretical predictions on the ground state masses \cite {Giannini2001,Ebert2005},
there seemed to have little agreement among the model predictions of the properties
like spin-hyperfine spiltting among the $J^{P}=\frac{1} {2}^{+}$ and $J^{P}=\frac{3}
{2}^{+}$ baryonic states, the form factors \cite{Giannini2001}, magnetic moments etc
\cite{Bhavin2008}. All these reasons make the study of the heavy flavour spectroscopy
extremely rich and interesting.\\
\\ Recently, the SELEX Collaboration \cite{A. Ocherashvili2005} has reported the
discovery of $\Xi_{cc}^{++}$ with a mass of $3.519$ Ge$V$. However, other groups such
as BABAR \cite{B. Aubert2006}, BELLE \cite{R. Chistov2006} and FOCUS \cite{S. P.
Ratti2003} have all failed to confirm this state. So there is a demand to review this
state both theoretically and experimentally. Also, there is renewed interest in the
static properties of heavy flavour baryons such as its masses and magnetic
moments\cite{Santopinto E1998,Ebert D2005,Faessler A2006,Athar S B2005,De La
Cruz2006,Bali G S2001,Avery P1995,Mattson M2002,Patel B2006}. Baryons containing two
heavy quarks are important and interesting systems to study the quark-diquark structure
of baryons and to understand the dynamics of light quark in the vicinity of heavy
quarks. Many of these narrow hadron resonances expected to be observed experimentally,
might bring many surprises in QCD spectroscopy\cite{Rosner609}. \\
\\We report
here the mass spectra and magnetic moments of $ccq(q\in u,d,s)$ systems using coulomb
plus martin potential within the quark-diquark model of the Baryon. Here the two heavy
quarks are considered for the diquark
states.\\

\noindent {\large{\bf Theoretical framework}}\\
\\ The notion of diquark is as old as the quark model itself.
Gell-Mann \cite{M. Gell-Mann1964} mentioned the possibility of diquarks in his original
paper on quarks. Soon afterwards, Ida and Kobayashi\cite{M. Ida1966} and Lichtenberg
and Tassie \cite{D. B. Lichtenberg1967} introduced effective degrees of freedom of
diquarks in order to describe baryons as composed of a constituent diquark and quark.
Since its introduction, many articles have been written on this subject \cite{M.
Anselmino1993} up to the most recent ones \cite{F.Wilczek2005}. The presence of a
coherent diquark structure within baryons simplifies baryonic calculations by
considering diquark as antiquark. By treating diquark as antiquark, the problem of
three-body quark excitations within baryons reduces to that of a two-body
quark-diquark interaction. In the Present Study of doubly charm baryons, the two
 charm quarks are considered as the diquark states. \\
\\ The Hamiltonian of the baryon, in the diquark model, can be written in terms of
diquark Hamiltonian plus qurk-diquark Hamiltonian as \cite{W. S. Carvalho1994}
\begin{equation}
H=H_{jk}+H_{i,jk}
\end{equation}\label{eq:1}
The internal motion of the diquark($jk$) is described by
\begin{equation}
H_{d}=H_{jk}=\frac{{p}^2}{2m_{jk}}+V_{jk}(r_{jk})
\end{equation}
where, $p$ is the relative momentum of the quarks within the diquark.\\ The Hamiltonian
of the relative motion of the diquark($jk$) and the third quark($i$) is
\begin{equation}
H_{i,d}=H_{i,jk}=\frac{{q}^2}{2m_{i,jk}}+V_{i,jk}(r_{id})
\end{equation}\\
Here,\begin{equation} m_{jk}=\frac{m_{j} m_{k}}{m_{j}+m_{k}},
m_{i,jk}=\frac{m_{i}(m_{j}+ m_{k})}{m_{1}+m_{2}+m_{3}} \end{equation}

The diquark potential can be written as,
\begin{equation}\label{eq:05}
V_{jk}=-\frac{2}{3}\alpha_s\frac{1}{r_{jk}}+ b{\,\,}  r^{\nu}_{jk}
\end{equation}
and the quark-diquark potential
\begin{equation}
V_{i,jk}=-\frac{4}{3}\alpha_s\frac{1}{r_{id}}+ b{\,\,}  r^{\nu}_{id}
\end{equation}
where, $r_{id}$ is the quark-diquark separation distance, $\nu $ is a general power
corresponding to the confining part of the potential, such that $\nu =0.1$ in Eqn
(\ref{eq:05})
 represents  the coulomb plus Martin potential, $b$ is model parameter corresponding to the confining part
of the potential, which is assumed to be same for the di-quark interaction as well as
between the quark-diquark interaction. The running strong coupling constant
($\alpha_s(\mu)$) is computed using the relation
\begin{equation}
\alpha_s(\mu)=\frac{\alpha_s(\mu_0)}{1+\frac{33-2\,n_f}{12\,\pi}\alpha_s(\mu_0)ln(\frac{\mu}{\mu_0})}
\end{equation}
where, $\alpha_s(\mu_{0}=1 GeV)\approx 0.7$ has been used \cite{Ajay2005}.\\ For the
present study, we employ the variational method
and the hydrogenic wavefunction given by \\
\begin{equation}
 R_{nl}(r) = \left(\frac{\mu^3 (n-l-1)! }{2n(n+l)!}\right)^{1/2}(\mu \ r)^l  e^{-
\mu \ r /2} L^{2l+1}_{n-l-1}(\mu \ r)
\end{equation}\\
has been used as the trial wavefunction. Here, $n$, $l$ are the quantum numbers, $\mu$
is variational parameter and $L^{2l+1}_{n-l-1}(\mu r)$ is Laguerre polynomial. The
expectation value of the Hamiltonian described by Eqn(\ref{eq:toteg}) provides the
binding energy of the baryon
as\\
\begin{equation} \label{eq:toteg}E(\mu )=\left<H_{jk}\right>+\left<H_{i,jk}\right>
\end{equation}

\begin{table} \caption{The quark model parameters} \vspace{0.01in}
\begin{center}
\label{tab:01}
\begin{tabular}{llll}
\hline\hline Quark Masses
&$m_{u}=0.322$ (in GeV)\\
&$m_{d}=0.336$ (in GeV)\\
&$m_{s}=0.510$ (in GeV)\\
&$m_{c}=1.422$(in GeV)\\\\

Model Parameter&$b$ =0.197 Ge$V^{(\nu+1)}$\\
 \hline \hline
\end{tabular}
\end{center}
\end{table}

 The spin average mass of baryonic system (ie. without spin contribution) is then obtained as
\begin{equation}
M_{QQq}=\Sigma {m}_{i}+E(\bar\mu)\end{equation} The quark mass parameters and the
potential parameters of the model employed in our calculations are listed in Table
\ref{tab:01}. The spin average masses of $\Xi_{cc}^{++}$, $\Xi_{cc}^{+}$ and
$\Omega_{cc}^{+}$ baryons for different quark-diquark states are also listed in Table
\ref{tab:02}. \\
\\The degeneracy of the states are removed by introducing the spin dependent
interaction potential among the diquark($d$) as well as among the light
quark($l$)-diquark($l-d$) system given by \cite{S. S. Gershtein2000} \begin{eqnarray}
V^{(d)}_{SD}(r_{jk}) &=& \frac{1}{2}\ (\ \frac{L_{d} \ \cdot S_{d}}{2 m^{2}_c}
)(-\frac{dV(r_{jk})}{r_{jk} dr_{jk}} +\frac{8}{3}\alpha_s\frac{1}{r_{jk}^3}) \cr&& +
\frac{2}{3}\alpha_s \frac{1}{m^2_{c}}\frac{L_{d} \ \cdot S_{d}}{r_{jk}^3} \cr&& +
\frac{4}{3}\alpha_s
\frac{1}{3 m^2_{c}}{S_{c1} \ \cdot S_{c2}} [4\pi \delta(r_{jk})]\end{eqnarray}\\
for the diquark states, and
\begin{eqnarray}\label{eq:2.10}
V^{(l)}_{SD}(r)&=& \frac{1}{2}\ (\ \frac{L \ \cdot S_{d}}{2 m^{2}_c} +\frac{2
L.S_{l}}{2 m^2_{l}})(-\frac{dV(r)}{r dr}+\frac{8}{3}\alpha_s\frac{1}{r^3})\cr&&
+\frac{1}{3}\alpha_s \frac{1}{m_{c} m_{l}}\frac{L \ \cdot S_{d}+2 L\ \cdot S_{l}}{r^3}
\cr &&+\frac{4}{3}\alpha_s \frac{1}{3 m^2_{c}}({S_{d}+ L_{d}) \ \cdot S_{l}} [4\pi
\delta(r)]
\end{eqnarray}
for the ($l-d$) system. Where, $r$ is the relative co-ordinate of $l-d$ $(i,jk)$, $L$
is the relative angular momentum of the $l-d$ system and $S_{l}$ and $S_{d}$ are the
light quark and diquark spins, respectively. The first term in both expressions takes
into account the relativistic corrections to the potential $V(r)$. The second and third
terms are the relativistic corrections coming from the one-gluon exchange between the
quarks. The computed masses of $\Xi_{cc}^{++}$, $\Xi_{cc}^{+}$ and $\Omega_{cc}^{+}$
Baryons with different combinations of the quark-diquark states are listed in Table \ref{tab:03}.\\

\begin{table}\caption{Spin average masses of $\Xi_{cc}^{++}$, $\Xi_{cc}^{+}$ and
                 $\Omega_{cc}^{+}$ baryons.} \vspace{0.01in}
\begin{center}
\label{tab:02}
\begin{tabular}{llll}

\hline \hline
$n_{d}$(diquark)-&{\,\,\,\,\,\,\,\,}Present &Others&\\
$n_{l}$(light-quark)&{\,\,\,\,\,\,\,\,} &\cite{S. S. Gershtein2000}&\cite{Ebert2005}\\
 \hline
 \hline

1S 1s&{\,\,\,\,\,\,\,\,\,\,\,}3.544  &3.560&3.673\\
1P 1s&{\,\,\,\,\,\,\,\,\,\,\,}3.614& 3.790&3.898 \\
2S 1s&{\,\,\,\,\,\,\,\,\,\,\,}3.618 & 3.900&3.968 \\
1S 1p&{\,\,\,\,\,\,\,\,\,\,\,}3.648& 4.030& 4.077\\
1S 2s&{\,\,\,\,\,\,\,\,\,\,\,}3.652 &$-$& $-$\\
1P 1p&{\,\,\,\,\,\,\,\,\,\,\,}3.718 & 4.250&4.166 \\
1P 2s&{\,\,\,\,\,\,\,\,\,\,\,}3.722& $-$& $-$\\
2S 1p&{\,\,\,\,\,\,\,\,\,\,\,}3.722 & 4.360&$-$ \\

\hline

1S 1s&{\,\,\,\,\,\,\,\,\,\,\,}3.551  &3.560&3.673\\
1P 1s&{\,\,\,\,\,\,\,\,\,\,\,}3.621& 3.790&3.898 \\
2S 1s&{\,\,\,\,\,\,\,\,\,\,\,}3.625 & 3.900&3.968 \\
1S 1p&{\,\,\,\,\,\,\,\,\,\,\,}3.657& 4.030& 4.077\\
1S 2s&{\,\,\,\,\,\,\,\,\,\,\,}3.661 &$-$& $-$\\
1P 1p&{\,\,\,\,\,\,\,\,\,\,\,}3.727 & 4.250&4.166 \\
1P 2s&{\,\,\,\,\,\,\,\,\,\,\,}3.731 & $-$&$-$ \\
2S 1p&{\,\,\,\,\,\,\,\,\,\,\,}3.731 & 4.360&$-$ \\

\hline
1S 1s&{\,\,\,\,\,\,\,\,\,\,\,}3.675  &$-$&3.825\\
1P 1s&{\,\,\,\,\,\,\,\,\,\,\,}3.745& $-$&4.052 \\
2S 1s&{\,\,\,\,\,\,\,\,\,\,\,}3.749 & $-$& 4.124\\
1S 1p&{\,\,\,\,\,\,\,\,\,\,\,}3.788& $-$& \\
1S 2s&{\,\,\,\,\,\,\,\,\,\,\,}3.803 &$-$& \\
1P 1p&{\,\,\,\,\,\,\,\,\,\,\,}3.858 & $-$ \\
1P 2s&{\,\,\,\,\,\,\,\,\,\,\,}3.862 & $-$& \\
2S 1p&{\,\,\,\,\,\,\,\,\,\,\,}3.873 & $-$& \\

\hline
\end{tabular}
\end{center}
\end{table}

\begin{table}\caption{\label{tab:03}The mass spectrum of $\Xi_{cc}^{++}$, $\Xi_{cc}^{+}$ and
$\Omega_{cc}^{+}$ Baryons (Masses in GeV).}
\begin{center}
\begin{tabular}{llll}

\hline \hline
($n_{d}l_{d}n_{l}l$)$J^P$&{\,\,\,\,\,\,\,\,}Present &Others&\\
& &\cite{S. S. Gershtein2000}&\cite{Ebert2005}\\
\hline
 \hline
($\Xi_{cc}^{++}$) &  &  & \\
(1S 1s)$1/2^+$&{\,\,\,\,\,\,\,\,\,\,\,}3.519  &3.478&3.620\\
(1S 1s)$3/2^+$&{\,\,\,\,\,\,\,\,\,\,\,}3.555 & 3.610&3.727\\
(1P 1s)$1/2^-$&{\,\,\,\,\,\,\,\,\,\,\,}3.586& 3.702&3.838 \\
(2S 1s)$1/2^+$&{\,\,\,\,\,\,\,\,\,\,\,}3.596 & 3.812&3.910\\
(1S 1p)$1/2^-$&{\,\,\,\,\,\,\,\,\,\,\,}3.626 & 3.927&4.053\\
(1P 1s)$3/2^-$&{\,\,\,\,\,\,\,\,\,\,\,}3.628 & 3.834&3.959 \\
(2S 1s)$3/2^+$&{\,\,\,\,\,\,\,\,\,\,\,}3.629 & 3.944&4.027\\
(1S 2s)$1/2^+$&{\,\,\,\,\,\,\,\,\,\,\,}3.633 & $-$ &$-$\\
(1S 1p)$3/2^-$&{\,\,\,\,\,\,\,\,\,\,\,}3.635 & 4.039&4.101\\
(1S 1p)$1/2^{'-}$&{\,\,\,\,\,\,\,\,\,\,\,}3.639& 4.052&4.136\\
(1S 1p)$3/2^{'-}$&{\,\,\,\,\,\,\,\,\,\,\,}3.648 & 4.034&4.196\\
(1S 1p)$5/2^{'-}$&{\,\,\,\,\,\,\,\,\,\,\,}3.659 & 4.047&4.155\\
(1S 2s)$3/2^+$&{\,\,\,\,\,\,\,\,\,\,\,}3.660 & $-$ &$-$\\
(1P 2s)$1/2^-$&{\,\,\,\,\,\,\,\,\,\,\,}3.700 & $-$ &$-$\\
(1P 2s)$3/2^-$&{\,\,\,\,\,\,\,\,\,\,\,}3.733 & $-$&$-$\\

\hline
($\Xi_{cc}^{+}$) &  &   \\
(1S 1s)$1/2^+$&{\,\,\,\,\,\,\,\,\,\,\,}3.527  &3.478&3.620\\
(1S 1s)$3/2^+$&{\,\,\,\,\,\,\,\,\,\,\,}3.562 & 3.610&3.727\\
(1P 1s)$1/2^-$&{\,\,\,\,\,\,\,\,\,\,\,}3.594& 3.702 &3.838 \\
(2S 1s)$1/2^+$&{\,\,\,\,\,\,\,\,\,\,\,}3.604 & 3.812&3.910\\
(1P 1s)$3/2^-$&{\,\,\,\,\,\,\,\,\,\,\,}3.635 & 3.834&3.959 \\
(2S 1s)$3/2^+$&{\,\,\,\,\,\,\,\,\,\,\,}3.636 & 3.944&4.027\\
(1S 1p)$1/2^-$&{\,\,\,\,\,\,\,\,\,\,\,}3.639&3.927&4.053 \\
(1S 2s)$1/2^+$&{\,\,\,\,\,\,\,\,\,\,\,}3.642 & $-$&$-$ \\
(1S 1p)$3/2^-$&{\,\,\,\,\,\,\,\,\,\,\,}3.644 & 4.039&4.101 \\
(1S 1p)$1/2^{'-}$&{\,\,\,\,\,\,\,\,\,\,\,}3.648 & 4.052&4.136\\
(1S 2s)$3/2^+$&{\,\,\,\,\,\,\,\,\,\,\,}3.669 & $-$&$-$ \\
(1P 2s)$1/2^-$&{\,\,\,\,\,\,\,\,\,\,\,}3.709 & $-$&$-$ \\
(1S 1p)$3/2^{'-}$&{\,\,\,\,\,\,\,\,\,\,\,}3.710 &4.034&4.196\\
(1S 1p)$5/2^{'-}$&{\,\,\,\,\,\,\,\,\,\,\,}3.723 & 4.047&4.155\\
(1P 2s)$3/2^-$&{\,\,\,\,\,\,\,\,\,\,\,}3.742 & $-$&$-$\\

\hline
\end{tabular}
\end{center}
\end{table}

\begin{table}
\begin{center}
\begin{tabular}{lllll}
\hline \hline
($n_{d}l_{d}n_{l}l$)$J^P$&{\,\,\,\,\,\,\,\,}Present &Others&\\
& &\cite{S. S. Gershtein2000}&\cite{Ebert2005}\\
\hline \hline
($\Omega_{cc}^{+}$) &  & \\
(1S 1s)$1/2^+$&{\,\,\,\,\,\,\,\,\,\,\,}3.648  &$-$&3.778\\
(1S 1s)$3/2^+$&{\,\,\,\,\,\,\,\,\,\,\,}3.688 & $-$&3.872\\
(1P 1s)$1/2^-$&{\,\,\,\,\,\,\,\,\,\,\,}3.715& $-$ &4.002 \\
(2S 1s)$1/2^+$&{\,\,\,\,\,\,\,\,\,\,\,}3.725 & $-$&4.075\\
(1P 1s)$3/2^-$&{\,\,\,\,\,\,\,\,\,\,\,}3.761 & $-$&4.102 \\
(2S 1s)$3/2^+$&{\,\,\,\,\,\,\,\,\,\,\,}3.762 & $-$&4.174\\
(1S 1p)$1/2^-$&{\,\,\,\,\,\,\,\,\,\,\,}3.772 & $-$&4.208\\
(1S 1p)$3/2^-$&{\,\,\,\,\,\,\,\,\,\,\,}3.777 & $-$&4.252\\
(1S 1p)$1/2^{'-}$&{\,\,\,\,\,\,\,\,\,\,\,}3.781 & $-$&4.271\\
(1S 2s)$1/2^+$&{\,\,\,\,\,\,\,\,\,\,\,}3.785 & $-$&$-$ \\
(1S 1p)$3/2^{'-}$&{\,\,\,\,\,\,\,\,\,\,\,}3.787 & $-$&4.303\\
(1S 1p)$5/2^{'-}$&{\,\,\,\,\,\,\,\,\,\,\,}3.795 & $-$&4.325\\
(1S 2s)$3/2^+$&{\,\,\,\,\,\,\,\,\,\,\,}3.811 & $-$&$-$ \\
(1P 2s)$1/2^-$&{\,\,\,\,\,\,\,\,\,\,\,}3.852 & $-$&$-$ \\
(1P 2s)$3/2^-$&{\,\,\,\,\,\,\,\,\,\,\,}3.884 & $-$&$-$\\
\hline
\end{tabular}
\end{center}
\end{table}

\begin{table}\caption{\label{tab:04} Magnetic moments  of the $\Xi_{cc}^{++}$, $\Xi_{cc}^{+}$ and
$\Omega_{cc}^{+}$ Baryons in terms of nuclear magneton $\mu_{N}$.}
\begin{center}
\begin{tabular}{lllllll}

\hline
\hline

(1S1s)&Present& H.M\cite{Bhavin2008} &RQM \cite{Amand2006}&NRQM \cite{Amand2006}&AL1\cite{Albertus2006} \\
\hline \hline
($\Xi_{cc}^{++}$)  &-0.054&-0.137&0.130&-0.010&-0.208\\
($\Xi_{cc}^{*++}$) &2.513&2.749&$-$&$-$&2.670\\
($\Xi_{cc}^{+}$)   &0.808&0.859&0.720 &0.740& 0.785\\
($\Xi_{cc}^{*+}$)  &-0.045&-0.168&$-$&$-$&-0.311\\
($\Omega_{cc}^{+}$)&0.727&0.783&0.670& 0.670& 0.635\\
($\Omega_{cc}^{*+}$)&0.242&0.121&$-$&$-$&0.139\\

 \hline

\end{tabular}
\end{center}
\end{table}

\noindent {\large{\bf Magnetic Moments of the Double Heavy Flavour Baryons}}\\
\\For the computation of the magnetic moments, we consider the mass of bound quarks
inside the baryons as its effective mass taking in to account of its binding
interactions with other two quarks. The effective mass for each of the constituting
quark  $m^{eff}_{i}$ can be defined as \cite{Bhavin2008}
\begin{equation}
m^{eff}_{i}=m_i\left( 1+\frac{E(\bar\mu)}{\sum\limits_{i}m_{i}}\right)
\end{equation}
such that the corresponding mass of the baryon is given by
\begin{equation}
M_B=\sum\limits_{i}m_{i}+E(\bar\mu)=\sum\limits_{i}m^{eff}_{i}\\
\end{equation}
Now, the magnetic moment of baryons are obtained in terms of the bound quarks as
\cite{Bhavin2008}

\begin{equation}
\mu_B=\sum\limits_{i}\left<\phi_{sf}\mid\mu_{i}
\overrightarrow{\sigma}_{i}\mid\phi_{sf}\right>
\end{equation}
where
\begin{equation}
\mu_{i}=\frac{e_{i}}{2m_{i}^{eff}}
\end{equation}
Here, $e_{i}$ and $\sigma_{i}$ represents the charge and the spin of the quark
constituting the baryonic state.$\left|\phi_{sf}\right>$ represents the spin-flavour
wave function of the respective baryonic state \cite{Bhavin2008}. By using spin flavour
wave function corresponds to $J^{P}=\frac{1} {2}^{+}$ and $J^{P}=\frac{3} {2}^{+}$
\cite{Bhavin2008}, we compute the magnetic moments of the baryons containing double
charm quarks. Our results are listed in Table \ref{tab:04}. Other theoretical model
predictions of the magnetic moments are also listed for comparison.\\

\noindent {\large{\bf Results and Discussions}}\\

We have employed a simple nonrelativistic approach with coulomb plus martin potential
to study the masses of the double charm baryons in the quark-diquark model. The model
parameters are obtained to get the ground state spin average masses of the ccq systems.
The mass spectra of $\Xi_{cc}^{++}$, $\Xi_{cc}^{+}$ and $\Omega_{cc}^{+}$ Baryons are
listed in Table \ref{tab:03}. The extra states predicted in this picture would require
experimental verification in support of the quark-diquark structure of the double heavy
flavour baryons. In conclusion, our results of doubly charmed baryons are
found to be in accordance with other model predictions.\\
\\ It is important to note that the predictions of the magnetic moment of double heavy
Baryons studied here are with no additional parameters. Our results on magnetic moments
are compared with one of our recent predictions using hypercentral potential
\cite{Bhavin2008} as well as with relativistic (RQM) and the nonrelativistic quark
model (NRQM) predictions of \cite{Amand2006} in Table \ref{tab:04}. The special feature
of the present study to compute the magnetic moments of double heavy flavour baryons is
the consideration of the effective interactions of the bound state quarks by defining
an effective bound state mass to the quarks within the baryon, which vary
according to different interquark potential as well as with quark compositions.\\
\\
Experimental measurements of the heavy flavour baryon magnetic moments are difficult
and only few experimental groups (BTeV and SELEX Collaborations) are expected to do
measurements in near future. \\

\noindent {\large{\bf Acknowledgments}}\\
\\We acknowledge the financial support from University Grant Commission, Government of
India, under a Major Research Project \textbf{F. 32-31/2006(SR)}.\\

\end{document}